\documentclass[10pt,journal,twocolumn,twoside]{IEEEtran} 
\usepackage{graphicx}
\usepackage{epstopdf}
\usepackage{float}
\usepackage{algorithmic}
\usepackage{array}
\usepackage{amsmath}
\usepackage{amssymb}
\usepackage{mdwmath}
\usepackage{eqparbox}
\usepackage{stfloats}
\usepackage{hyperref}
\usepackage{cleveref}
\hypersetup{hypertex=true,
            colorlinks=true,
            linkcolor=blue,
            anchorcolor=blue,
            citecolor=blue}
\usepackage{cases} 
\usepackage{subfigure}
\usepackage{upgreek}
\usepackage{makecell}
\usepackage{amsmath}
\usepackage[boxed,ruled,commentsnumbered]{algorithm2e}
\usepackage{url}
\usepackage{cite}
\usepackage[table,xcdraw]{xcolor}
\usepackage{colortbl}
\ifCLASSOPTIONcompsoc
\usepackage[caption=false,font=normalsize,labelfont=sf,textfont=sf]{subfig}
\else
\allowdisplaybreaks[4]

\makeatletter

\renewcommand*{\@opargbegintheorem}[3]{\trivlist
      \item[\hskip \labelsep{\bfseries #1\ #2}] \textbf{(#3):}\ }
\makeatother

\begin{document}
\title{Semantic-Enabled 6G Communication: A Task-oriented and Privacy-preserving Perspective}
\author{Shuaishuai Guo,~\IEEEmembership{Senior Member, IEEE}, Anbang Zhang, Yanhu Wang, Chenyuan Feng,~\IEEEmembership{Member, IEEE}, \\ and Tony Q. S. Quek,~\IEEEmembership{Fellow, IEEE}
\thanks{Shuaishuai Guo, Anbang Zhang, and Yanhu Wang are with Shandong University, China; Chenyuan Feng (corresponding author) is with  EURECOM, France; Tony Q. S. Quek is with Singapore University of Technology and Design, Singapore.}
}
\maketitle
\begin{abstract} 
Task-oriented semantic communication (ToSC) emerges as an innovative approach in the 6G landscape, characterized by the transmission of only vital information that is directly pertinent to a specific task. While ToSC offers an efficient mode of communication, it concurrently raises concerns regarding privacy, as sophisticated adversaries might possess the capability to reconstruct the original data from the transmitted features. This paper provides an in-depth analysis of privacy-preserving strategies specifically designed for ToSC relying on deep neural network-based joint source and channel coding (DeepJSCC). Our study encompasses a detailed comparative assessment of trustworthy feature perturbation methods such as differential privacy (DP) and encryption, alongside intrinsic security incorporation approaches like adversarial learning to train the JSCC and learning-based vector quantization (LBVQ). Our comparative analysis underscores the integration of advanced explainable learning algorithms into communication systems, positing a new benchmark for privacy standards in the forthcoming 6G era.
\end{abstract}

\section{Introduction} 
The integration of artificial intelligence (AI) in 6G networks is anticipated to revolutionize industries by enabling new business models and services \cite{8869705}. The network's reliability, trustworthiness and timeliness will be critical in the scenario of massive mobile users with real-time response requirements \cite{9759241}. Despite the promising prospects, the practical implementation continues to encounter numerous unprecedented hurdles, particularly for burst communications. Task-oriented semantic communication (ToSC), emerging as a promising paradigm, is primarily characterized by its selective transmission of information \cite{10644029}. The fundamental concept of ToSC is to enhance communication efficiency and task performance by transmitting task-relevant semantic information, rather than raw data. This approach has garnered considerable attention, chiefly due to its proficiency in enhancing efficiency and reducing latency through the minimization of data transmission volume. For instance, a semantic communication approach with limited knowledge representation is proposed in \cite{10558819}, while a multimodal framework driven by AI models is introduced in \cite{10670195} to achieve low-latency. Furthermore, ToSC exhibits the capacity to offer a more customized and efficient user experience. The targeted and efficient nature of ToSC, therefore, not only aligns with the technological advancements envisaged in 6G communications but also caters to the nuanced demands of modern digital applications, ensuring a seamless and user-oriented interaction.

To alleviate the misunderstandings and incorrect interpretations, ToSC can increase information clarity, relevance, transparency, credibility, and verifiability by concentrating on task-relevant information, implementing well-designed mechanisms of channel coding and feedback.  Although this strategy inherently provides a certain level of privacy since selective data transmission eliminates the unnecessary data sharing, the information conveyed may still be vulnerable \cite{10483054}.  If intercepted, even these task-relevant data bits could reveal personal or sensitive information \cite{2020Adversarial}. This privacy-leakage risk is heightened by the advanced capabilities in machine learning (ML) and data analysis, which might enable adversaries to extract significant insights from minimal data.  Compared to a generic ToSC framework \cite{10644029} for various tasks with diverse data types, this paper has refined it for privacy and security applications. First, we discuss ongoing research aimed at developing privacy-preserving methods specifically for ToSC relying on deep neural network-based joint source and channel coding (DeepJSCC). Second, we aim to devise techniques that can be seamlessly incorporated into ToSC without significantly compromising their efficacy and efficiency, or in other words, strike  a balance between utility, efficiency, and privacy. At last, we evaluate and contrast feature perturbation methodologies, such as differential privacy (DP) and encryption techniques, with intrinsic security incorporation approaches like adversarial learning and learning-based vector quantization (LBVQ). Our analysis highlights the potential for integrating sophisticated learning algorithms into
contemporary communication systems.

\section{Shift From Task-Agnostic Communications to ToSC and Privacy Challenges}

 \begin{figure*}[htbp]
\centering
\includegraphics[width=2\columnwidth]{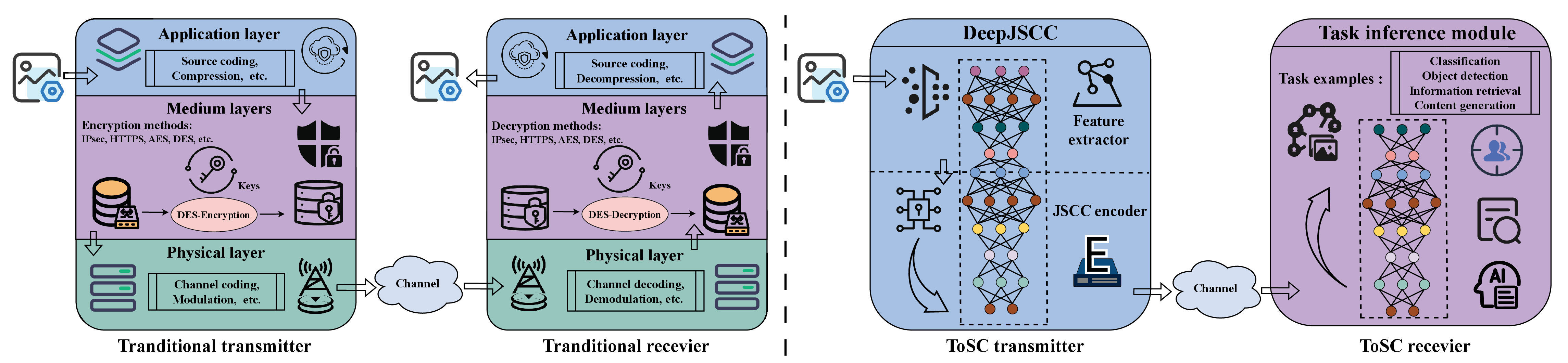}
\caption{Traditional transceivers versus ToSC transceivers.}
\label{fig1}
\end{figure*}

Within the traditional framework of source-channel separation, the identification, representation, and transmission of information are rigorously addressed by rate-distortion theory and channel coding theory, respectively. This paradigm, which prioritizes reconstruction-oriented compression and task-agnostic communication, has underpinned several iterations of digital communication systems. However, the advent of machine-to-machine communications and human-machine interactions necessitates a reassessment of this paradigm, considering that exact reconstructions are often of secondary importance from a machine's perspective. Notably, task-specific descriptors, derived via ML algorithms from latent feature spaces, are substantially more concise than their counterparts used for reconstruction purposes. Furthermore, communication systems trained end-to-end can significantly surpass those designed based on source-channel separation, across various performance metrics.

In ToSC systems, data transmission is meticulously tailored to align with the receiver's requirements. As shown in Fig. \ref{fig1}, different from traditional transceivers, the ToSC system comprises two main components: the “DeepJSCC” and the “Task inference module”. The DeepJSCC contains a JSCC encoder responsible for task-related information extraction, compression, and protection (against both channel noise and adverseries' attacks). The output passes through a channel to the task inference module, which can handle tasks such as classification, object detection, information retrieval, and content generation. Inside, it displays a multi-layered network topology that indicates intricate processing is occurring to perform the given tasks.

Despite these advantages, ToSC introduces significant privacy concerns. The main issue stems from the nature of the information being transmitted. Although ToSC systems transmit only task-specific information, this data can be sensitive and vulnerable to privacy infringements. Given the unpredictability of adversaries' objectives, it is imperative to devise a comprehensive and efficacious strategy for safeguarding a spectrum of private data. For instance, in the transmission of facial images, adversaries may undertake diverse strategies to extract personal attributes such as gender or skin tone, or alternatively, engage in face recognition. In our research, we postulate the quality of image reconstruction—evaluated by metrics such as mutual information (MI) leakage—as a quantifiable metric for privacy preservation. Our approach assumes that amplifying distortions in data reconstructed by potential intruders indirectly shields various aspects of personal information. This approach resonates with the principles of perturbative privacy preservation, notably exemplified by DP paradigms. By escalating the degree of distortion, we can significantly diminish the likelihood or amplify the challenge for adversaries in gleaning sensitive information, thereby strengthening the robustness of privacy protection.

\section{Privacy Preservation Methods for ToSC}
Most ToSC systems are based on DeepJSCC architectures and employ an end-to-end training methodology to extract high-dimensional task-related channel-robust features for transmission. This approach ensures a coherent and automatically optimized process from data input to the final task output. Many traditional privacy protection techniques, such as $k$-anonymity,  $l$-diversity, and $t$-closeness,  are often designed to protect user privacy by modifying data and are not suitable for working with complex or high-dimensional data \cite{2020Adversarial}, as the process of making records indistinguishable can lead to significant data loss or impracticality in datasets with numerous attributes. A thorough analysis of privacy-preserving strategies specifically designed for ToSC is provided, along with a comparative assessment of feature perturbation methods and intrinsic security incorporation approaches.

\subsection{Feature Perturbation Methods}
Conventional approaches concentrate on the safeguarding of original data, typically through direct pre-processing under the assumption that the entire AI model is in the possession of a potentially untrustworthy third party. However, in DeepJSCC-based ToSC, the encoder belongs to the transmitter, which plays the role of task-related information extraction, compression, and protection.  The employment of DeepJSCC makes conventional techniques applied in-between source and channel coding difficult in DeepJSCC. Applying these techniques will prevent the DeepJSCC to extract the task-related and channel-robust features, thus the only feasible is to process the output of DeepJSCC. Therefore, traditional data perturbations should be replaced by the feature perturbations on the DeepJSCC output. Next, we discuss two typical feature perturbation methods for privacy preservation.
\begin{itemize}
    \item DP provides a mathematical framework to quantify privacy loss and offers strong guarantees by adding noise to DeepJSCC outputs. Its robustness and theoretical assurances make it widely adopted. However, the noise added to protect individual data points can reduce data utility, particularly in scenarios requiring precise semantic information for task inference. While excessive noise enhances privacy, it inherently degrades task inference performance, impacting the utility of transmitted data.
    \item Encryption serves as a fundamental method for securing DeepJSCC outputs by making them unintelligible without a decryption key. However, it can slow down training and inference. Without channel coding, encryption is also vulnerable, potentially compromising decryption integrity. To address these issues, simple or error-robust encryption methods, like key-guided feature shuffling \cite{10107616}, are recommended for ToSC. Additionally, encryption-based methods face key-sharing burdens and the key leakage.
    \end{itemize}

\begin{figure*}[htbp]
\centering
\includegraphics[width=2.1\columnwidth]{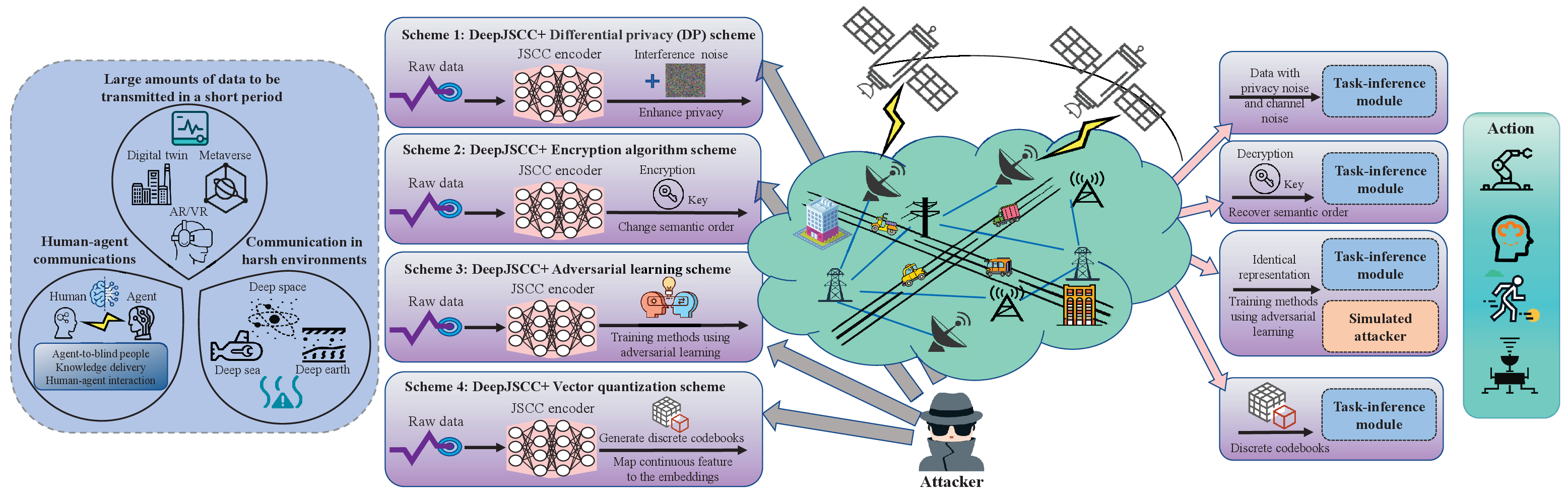}
\caption{Transceiver structure of four privacy protection methods: differential privacy, encryption, adversarial learning, and vector quantization.}
\label{fig2}
\end{figure*}

\subsection{Intrinsic Security Incorporation Strategies}
Another approach involves designing DeepJSCC with intrinsic security. Two typical strategies are as follows.
\begin{itemize}
    \item Adversarial learning trains models in a competitive setup involving a generator and a discriminator. The generator creates data indistinguishable from real data, while the discriminator aims to differentiate between fake and real data, enhancing both models' performance. This technique is vital for ToSC privacy protection. It generates feature representations useful for tasks yet hard for adversaries to exploit, crucial in protecting sensitive data. The adversarial process preserves data's essential characteristics while minimizing an adversary's ability to extract sensitive information, balancing utility and privacy. Notably, the transmitter is unaware of the adversary's neural network architecture, which is impractical to obtain. A simulated adversary can be constructed and used in adversarial learning to assist ToSC training \cite{10458014}, defending against model inversion attacks (MIAs). MIAs, a common privacy threat to ML models, involve adversaries using access to trained models to reconstruct sensitive training data. Even without knowing the adversary's inversion network architecture, this approach effectively protects privacy through a simulated adversary.
    \item LBVQ uses discrete latent representations which are more robust against inversion. This means that it's harder to reconstruct the original input data accurately from the latent representations. The quantization step in LBVQ, which converts continuous latent variables into discrete ones, acts as a bottleneck, reducing the amount of detailed information that can be decoded from the latent space. This inherent characteristic of LBVQ makes it a suitable choice for tasks where preserving privacy is crucial. The integration of LBVQ within ToSC systems presents an additional benefit regarding compatibility with existing digital communication frameworks. In current ToSC models, features extracted by neural networks are typically continuous, aligning well with analog communications but not with prevalent digital communication systems. Fortunately, LBVQ produces discrete representations that can be directly mapped to digital modulation symbols. This compatibility is instrumental in facilitating a seamless transition from well-established digital communication systems to the emerging ToSC paradigms.
\end{itemize}

\begin{table*}[]
\centering
\setlength{\tabcolsep}{11pt}
\caption{Computation complexity, model training cost and learning latency of 4 privacy-preserving mechanisms over CIFAR-10 dataset.}\label{tab:1}
\renewcommand{\arraystretch}{1.4}
\scalebox{1.1}{\begin{tabular}{|
>{\columncolor[HTML]{dddfff}}c |
>{\columncolor[HTML]{dddfff}}c |
>{\columncolor[HTML]{E8DAEF}}c|
>{\columncolor[HTML]{dddfff}}c|
>{\columncolor[HTML]{E8DAEF}}c |
>{\columncolor[HTML]{dddfff}}c|}
\hline
                      \cellcolor[HTML]{BB8FCE}    & 
                      \cellcolor[HTML]{BB8FCE}DeepJSCC    & \cellcolor[HTML]{BB8FCE}DeepJSCC-DP & \cellcolor[HTML]{BB8FCE}DeepJSCC-Encryption & \cellcolor[HTML]{BB8FCE}IBAL & \cellcolor[HTML]{BB8FCE}DeepJSCC-LBVQ \\ \hline
FLOPs & 0.085 G               & 0.085 G                                & 0.085 G               & 0.382 G                        & 0.477 G        \\ \hline
Params   & 3.19 M              & 3.19 M                           & 3.19 M                & 10.91 M                        & 12.61 M         \\ \hline
Train Time for 1 Epoch & 8.23 s  & 8.93 s                            & 8.79 s               & 32.5 s                        & 37.30 s        \\ \hline
Test Time for 1 Instance   & 0.002 s     & 0.002 s                          & 0.002 s              & 0.002 s                        & 0.004 s       \\ \hline
\end{tabular}}
\end{table*}

\subsection{Transceiver Structure Comparison}
Comparing feature perturbation methods and encryption with intrinsic security incorporation strategies reveals distinct strengths and weaknesses, as demonstrated in Fig. \ref{fig2}. Feature perturbation methods like DP and encryption offer robust theoretical guarantees for privacy. DP provides a quantifiable measure of privacy by adding noise to the features. Encryption provides strong security for DeepJSCC output in transit but does not address the unique challenges of vulnerability to channel errors and real-time processing.  While adversarial learning offers an intrinsic security incorporation approach in ToSC. By introducing the adversary loss in the training phase of DeepJSCC, attacking-robust communication models can be obtained. Adversarial learning maintains strong performance even in the face of unknown or intentional attacks. However, the complexity and computational demands of adversarial models are notable challenges in the training phase. Additionally, continuous updates and model improvements may be needed to cope with new attack methods. LBVQ offers an effective approach for safeguarding feature privacy through the map of features into smaller vector spaces. This method significantly reduces feature dimensions while upholding the integrity of information. Nevertheless, the process of dimension reduction might result in potential information loss. And another challenge is the need for appropriate vector quantization based on the specific tasks and data types involved.

\subsection{Cost and Delay Comparison}
Table \ref{tab:1} shows the computational cost, complexity, and communication latency of four privacy schemes running on an 11th Gen Intel(R) processor at 2.50 GHz and a single 3060 CPU core, including the specific number of floating-point operations (FLOPs) in the whole training process, the number of parameters in all the neural network model, the time for a single training epoch (batch-size: 512) and the task-inference time for a single image. The results indicate that DeepJSCC, DeepJSCC-DP, and DeepJSCC-Encryption demonstrate similar computational requirements and maintain a lower complexity profile, as evidenced by their FLOPs and parameter counts. In contrast, IBAL and DeepJSCC-LBVQ necessitate substantially greater resources, manifesting in increased FLOPs, a higher number of parameters, and extended training durations. Despite this, the inference times for all configurations, with the exception of DeepJSCC-LBVQ, remain comparably low. This suggests that all configurations, barring the LBVQ variant, achieve high efficiency during the inference phase. The LBVQ variant, however, incurs additional delays due to its discrete codebook mapping and remapping processes. Importantly, when compared to the baseline DeepJSCC model without privacy enhancements, the DeepJSCC-DP, DeepJSCC-Encryption, and IBAL models exhibit negligible increases in task-inference time, rendering them particularly suitable for applications requiring low-latency and privacy-sensitive remote inference.

\section{Experiments and Discussions}
\subsection{Experimental Settings}
\emph{1) Dataset and Attack Setups:} CIFAR-10 dataset and CelebA dataset are adopted for image classification task and face recognition task, respectively. The former comprises $60,000$ color images and categorized into $10$ distinct classes. The latter contains over 200,000 celebrity images with 40 attribute annotations per image, whose images cover a rich range of human pose variations and diverse background information. All networks are configured to produce outputs of identical dimensions. Furthermore, we incorporate a hypothetical scenario involving an adversarial attack network. This network is designed to execute model inversion via a black-box attack approach. It is posited to have continuous access to the network model on the target device, thereby enabling it to attempt image reconstruction. This scenario is pivotal in assessing the robustness of our framework against potential security breaches.

\emph{2) Performance Metrics:} For image classification task, classification accuracy and MI leakage are employed. The former serves as an indicator of inference performance, with higher classification accuracy signifying more effective inference capabilities. The latter, is utilized to gauge the level of privacy protection. A lower MI leakage value in the reconstructed images indicates that its attacker steals the transmitted data, having less privacy leakage of the reconstructed image, which also reflects the stronger privacy preservation, as it indicates that the reconstructed image has a reduced sensitive data details, which prevents unauthorised interpretation. For face recognition task, the top-$1$ accuracy and the reconstructed image of the attacker are used. The former is a metric of recognition performance and is used to judge the task performance. The latter is used to measure the level of privacy protection.

\emph{3) Approaches for Evaluation:}
To make a comprehensive comparison, we select four state-of-the-art ToSC schemes, including:
\begin{itemize}
    \item \textbf{DeepJSCC-DP:} DeepJSCC, originally designed for data-oriented communication systems, utilizes deep neural network-based encoders to map data directly to channel input symbols. Through the injection of Laplacian noise into the transmission characteristics \cite{9152658}, DP mechanism allows for precise control over the level of privacy by adjusting the privacy budget, which is set at $0.05$, $0.1$, and $0.9$ for our experiments. A pivotal aspect to note is that a smaller privacy budget correlates with a higher volume of noise injected into the transmitted features. This increased noise level consequently leads to stronger privacy protection, as it more effectively obscures the original data features, thereby enhancing the security against potential data breaches or unauthorized data reconstruction efforts.
    \item \textbf{DeepJSCC-Encryption:} In DeepJSCC-Encryption, the encoder not only processes the data to extract features for the JSCC, but also integrates an encryption operation into its output \cite{10107616}. This dual-functionality approach effectively combines feature extraction and encoding with a layer of cryptographic security. At the receiver, the process is reversed. The encoded and encrypted features are subjected to a decryption operation, a critical step for regaining the original data characteristics. Post-decryption, these features are then utilized for the intended classification and reconstruction tasks. This mechanism ensures that the data remains secure during transmission, only becoming accessible and usable upon successful decryption at the intended destination.
    \item \textbf{IBAL:} IBAL \cite{10458014} represents a novel scheme by leveraging the principles of adversarial learning. This method uniquely trains the encoder to effectively deceive the potential adversaries. It does so by optimizing the encoder to maximize the distortion in the data reconstruction process. Such a strategy is designed to thwart unauthorized attempts at data reconstruction, thereby enhancing the privacy and security of the transmitted information. 
    Moreover, variational feature encoding (VFE) \cite{Shao2021LearningTC} belongs to the theory based on information bottleneck (IB), which is similar to IBAL. The scheme incorporates its IB theory and we comparatively test the performance of strategies with IB and without adversarial learning in terms of utility as well as privacy preservation.
    \item \textbf{DeepJSCC-LBVQ:} DeepJSCC-LBVQ \cite{10159007} represents a sophisticated ToSC scheme that incorporates digital modulation. Its essence is the implementation of a robust encoder, which is underpinned by a learned codebook. Its primary objective is to enhance communication robustness in response to channel variations. The essence of DeepJSCC-LBVQ lies in its ability to effectively balance the trade-off between informativeness and robustness. By employing a learned codebook, the scheme adapts to varying channel conditions, ensuring that the integrity and reliability of the transmitted data are maintained, even in challenging communication environments. This adaptability makes it a significant contender in scenarios where channel variability is a critical factor. In addition, utility-informativeness-security-based ToSC (UIS-ToSC) \cite{10483054} employs a strategy that combines adversarial learning as well as VQ and can be attributed to the LBVQ group. This scheme efficiently solves the utility-informativeness-security trade-off issue inherent in ToSC systems.
\end{itemize}

The inference performance and the quality of image reconstruction are critically influenced by the dimensionality of the encoded representation. To facilitate fair comparisons, we have standardized the dimensionality of the representations encoded by all methods. For the continuous representation methods such as DeepJSCC-DP, DeepJSCC-Encryption, and IBAL, we employ a full-resolution constellation modulation technique. This approach is instrumental in maintaining the integrity and resolution of the encoded data during the modulation process. In contrast, for DeepJSCC-LBVQ, which is a discrete method, we utilize the $M$-ary quadrature amplitude modulation (QAM) scheme. This choice is tailored to suit the discrete nature of the representations encoded by DeepJSCC-LBVQ, ensuring that the modulation process is compatible with the encoding method. Additionally, to further ensure the impartiality of our evaluation, we have standardized the settings across all adversary attack networks. This uniformity is vital for ensuring that each method is subjected to non-discriminatory attacks, thereby providing a fair assessment of private level.

\subsection{Information Leakage Under Adversarial Attack}
Regarding the adversarial's description, we assume ToSC system is under black-box MIAs \cite{Fredrikson2015ModelIA}, where the adversaries reconstruct the received features as raw input using DNNs, and obtain users' privacy. Specifically, the adversary network is a DNN designed by the adversary and deployed on the adversary's device. And the transmitter's coded features are illegally accessed by the adversary. The adversary then attempts to generate an approximate reconstruction of the user's data based on the stolen transmission data. To improve the system's capability to combat attacks, we make a weak assumption that attacker knows the codebook and can continuously access the trained encoder. As the intentions of attacker are not known, we consider a universal loss function for training the attacker neural network, which is to minimize the distortion of reconstructed data. For image transmission task, the loss function for training attacker can be expressed as the sum of the average MSE loss and the average perception loss across all sampling images.  

\subsection{Experimental Results -- Image Classification}
Each scheme undergoes training at a specific signal-to-noise ratio (SNR), denoted as $\operatorname{SNR}_{\text {train}}$, which is set at $12$ dB. The testing phase involves varying SNR levels, specifically at $\operatorname{SNR}_{\text {test}}$ values of $4$ dB, $8$ dB, $12$ dB, $16$ dB, and $20$ dB. To maintain fairness, both discrete and continuous, the dimension of the encoded representation is uniformly set to $128$, which ensures that any observed differences in performance are attributable to the scheme's inherent characteristics rather than discrepancies in encoded representation size. Specifically, for DeepJSCC-LBVQ method, which employs a discrete approach, we use a codebook of size $16$. This size is selected as it offers a balance between complexity and performance. 

\begin{figure}
        \centering
        \setlength{\belowcaptionskip}{-0.3Cm}   
        \subfigure[Classification accuracy.]{{\label{fig3a}}\includegraphics[width=0.49\linewidth]{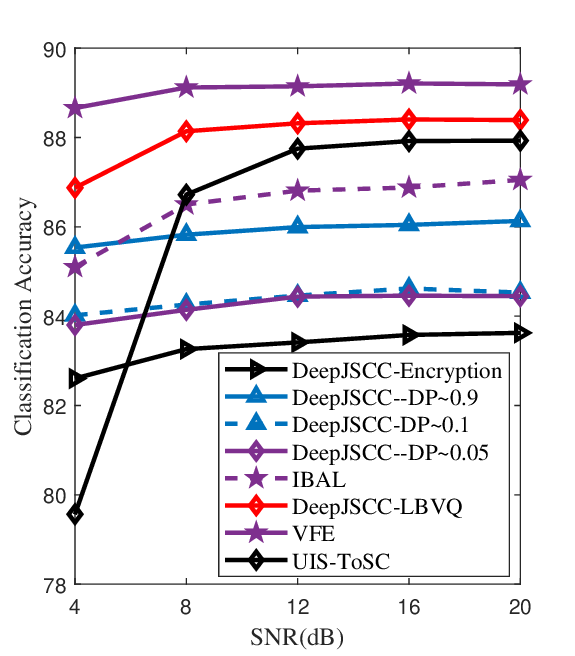}}
        \subfigure[MI Leakage.]{{\label{fig3b}}\includegraphics[width=0.49\linewidth]{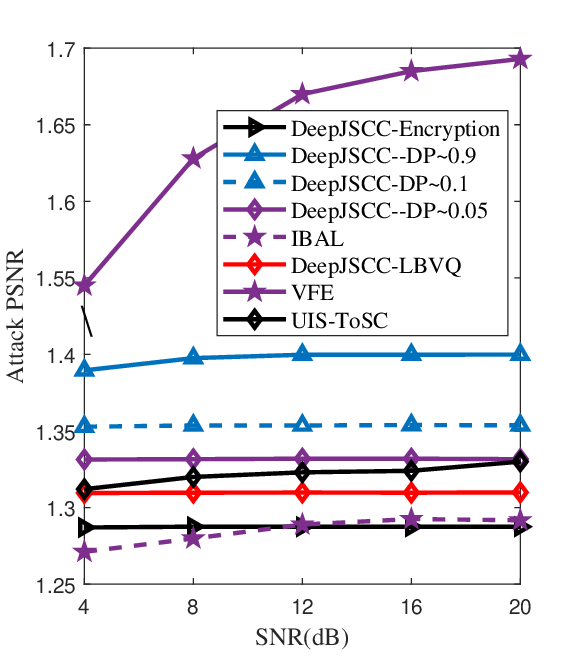}}
        \caption{Task-accomplishment Performance (i.e., image classification) versus Privacy-preserving Performance (i.e., MI Leakage) on the CIFAR-10 dataset.}
\end{figure}

The analysis of Figs. \ref{fig3a} and \ref{fig3b} reveals the significant insights into the image classification performance of various privacy-preserving schemes under an additive white Gassian noise (AWGN) channel using the CIFAR-10 dataset. First, the classification accuracies of IB group (IBAL and VFE) and VQ group (DeepJSCC-LBVQ and UIS-ToSC) are significantly higher under regular channel conditions (i.e. SNR $\ge$ 8dB) compared to DeepJSCC-DP and DeepJSCC-Encryption. 
Moreover, two type of schemes (VQ scheme and IB-based IBAL) also exhibit superior privacy protection capabilities than DeepJSCC-DP method. However, although the classification accuracy of VFE is excellent, the privacy preservation is slightly worse than our proposed scheme (MI $>$ 1.55). Under low SNR regimes (i.e. SNR $<$ 8dB), DeepJSCC-DP and DeepJSCC-Encryption show better robustness in terms of classification accuracy. At this point, the VQ group (DeepJSCC-LBVQ and UIS-ToSC) classification accuracy decays, due to the wide deflection of discrete features caused by the codebook indexing of the channel transmission, but still maintains good privacy preserving ability. And IBAL shows the best privacy-preserving capability, as it focuses on trade-off between both privacy and task, with a slight bias at low SNR.

Notably, DeepJSCC-DP, with its increased Laplacian noise injection, offers improved privacy protection at the cost of task performance. For instance, DeepJSCC-DP with a privacy budget of $0.05$ achieves similar privacy protection levels as DeepJSCC-LBVQ, but its classification accuracy falls behind by approximately $3$ to $4$ when SNR is greater than $8$dB. DeepJSCC-Encryption presents the best privacy protection among the compared methods. However, this comes at the expense of task performance, failing to strike an optimal balance between privacy and utility. This contrast highlights the superiority of the intrinsic security incorporation schemes (IBAL and DeepJSCC-LBVQ) over feature perturbation methods in achieving a well-balanced effect in both task performance and privacy preservation. The advanced IBAL scheme particularly stands out for its capability to improve privacy protection without significantly compromising task inference performance, achieving an optimal privacy-utility trade-off compared to the baseline methods. A key factor contributing to DeepJSCC-LBVQ's superior task performance is that its discrete representation, enhanced by the learned codebook, contains more informative messages, thereby leading to better performance. This underscores the effectiveness of discrete representation in ToSC systems.

\begin{figure*}
\centering
\includegraphics[width=1\linewidth]{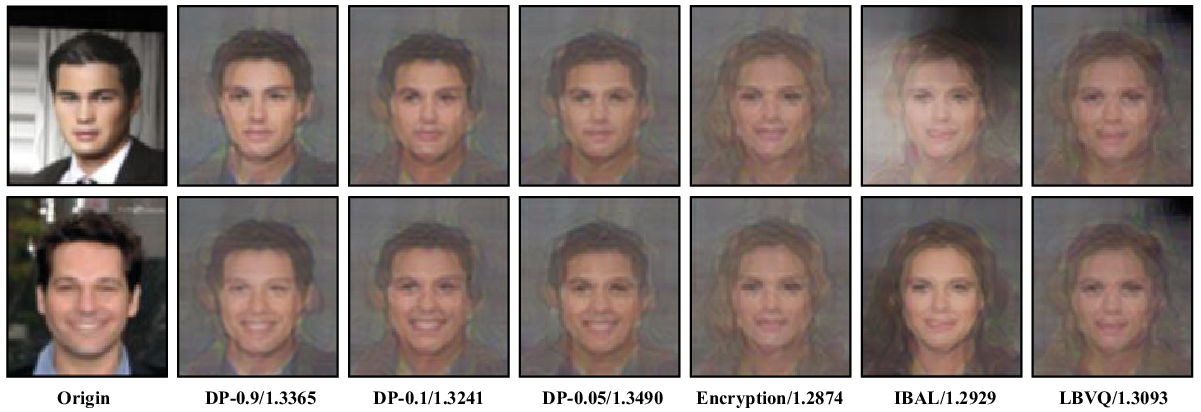}
\caption{Visual comparison about all schemes including DeepJSCC-DP-0.9, DeepJSCC-DP-0.1, DeepJSCC-DP-0.05, DeepJSCC-Encryption, IBAL, DeepJSCC-LBVQ (SNR$_{\text{train}}$=SNR$_{\text{test}}$=12dB) for the sample image of the CelebA dataset. The text below each image indicate the strategy and MI Leakage value.}
\label{fig4}
\end{figure*}

\begin{table}[]
\centering
\setlength{\tabcolsep}{2.5pt}
\caption{The Top-$1$ accuracy of face recognition in the CelebA dataset.}\label{tab:2}
\renewcommand{\arraystretch}{1.8}
\scalebox{1.15}{\begin{tabular}{|
>{\columncolor[HTML]{dddfff}}c |
>{\columncolor[HTML]{E8DAEF}}c|
>{\columncolor[HTML]{dddfff}}c|
>{\columncolor[HTML]{E8DAEF}}c|}
\hline
                      \cellcolor[HTML]{BB8FCE} & 
                      \cellcolor[HTML]{BB8FCE}Mustache& \cellcolor[HTML]{BB8FCE}Smiling & \cellcolor[HTML]{BB8FCE}Wavy Hair  \\ \hline
DP-0.05                 & $75.0\%\pm0.01\%$                              & $66.7\%\pm0.01\%$               & $72.9\%\pm0.01\%$                         \\ \hline
DP-0.1                 & $76.3\%\pm0.01\%$                              & $68.4\%\pm0.01\%$               & $73.6\%\pm0.00\%$                                 \\ \hline
DP-0.9                 & $76.5\%\pm0.01\%$                              & $67.2\%\pm0.01\%$               & $72.3\%\pm0.00\%$                              \\ \hline
Encryption                 & $77.6\%\pm0.00\%$                              & $67.2\%\pm0.00\%$               & $74.5\%\pm0.00\%$                                 \\ \hline
IBAL & $77.3\%\pm0.00\%$                              & $69.6\%\pm0.01\%$               & $71.8\%\pm0.01\%$                                 \\ \hline
LBVQ  &$85.3\%\pm0.00\%$                              & $79.9\%\pm0.01\%$               & $79.3\%\pm0.00\%$                               \\ \hline
\end{tabular}}
\end{table}

\subsection{Experimental Results -- Face Recognition}
Regarding the face recognition task, we consider a few single attributes as the target of the retrieval, e.g., smiling, moustache, wavy hair. Furthermore, the network requires significantly less information to predict a single characteristic than it does for $40$ attributes recognition, which should make it simpler to create privacy-protected features. To ensure fairness, each experiment was conducted five times. Table \ref{tab:2} presents the mean and standard deviation of Top-1 accuracy for face recognition under different strategies. As shown in Table \ref{tab:2}, DeepJSCC-LBVQ scheme achieves better task performance when performing face recognition with arbitrarily selected different attributes. This scheme, despite the fact that it undergoes dimensionality reduction and may lead to potential loss of information, adequately extracts the features needed for the face, maintains the integrity of the information and therefore ensures higher task performance. While IBAL, as a pre-trained model with adversarial learning, is also better able to extract the required features for face recognition task. Its performance under all three attributes is better than the DeepJSCC-DP scheme, but under some attributes (e.g., mustache, wavy hair) the performance is similar to the DeepJSCC-Encryption scheme. The reason is that IBAL considers the overall trade-off between performance and privacy, and outperforms both schemes in terms of privacy preservation, shown in Fig. \ref{fig4}. Longitudinally, the common DeepJSCC-DP and DeepJSCC-Encryption schemes have a greater loss (about 5$\%$-15$\%$ drop) in performance compared to the above DeepJSCC-LBVQ scheme. Moreover, the experimental results indicate that the standard deviation is remarkably small (less than 0.01\%), suggesting a stable training process and well-converged model. The minimal variance in loss and final outcomes across multiple training runs implies high consistency. Furthermore, this stability suggests that the model exhibits strong adaptability to the given task.

Next, we investigate the images of the CelebA dataset obtained after reconstructing the transmitted signals of these schemes using MIAs. As shown in Fig. \ref{fig4},  the attacker reconstructs the image worse with IBAL and DeepJSCC-LBVQ. IBAL even affects the attacker's judgment on the gender reconstruction of men and women, which brings great protection. Combined with the above analysis of performance, it can be seen that IBAL and DeepJSCC-LBVQ improve privacy protection while ensuring better task inference performance. As Laplacian noise increases, DeepJSCC-DP scheme improve their privacy protection, i.e., the attacker reconstructs a blurrier image. Additionally, the DeepJSCC-encryption scheme also provides good privacy protection because it completely disrupts the order of features in the image transmission, while the attacker only steals the image data and cannot carry out the complete reconstruction process later. Thus, IABL and DeepJSCC-LBVQ schemes achieve a better privacy-utility trade-off compared to other schemes.

\subsection{Scalability Issue}
The scalability of algorithms in large-scale networks is influenced by the number of devices, affecting design and performance. For DeepJSCC+DP, while more devices improve data utility, they also increase computational and communication costs. Dynamic privacy budget allocation and adaptive noise strategies are essential to address these challenges. Distributed differential privacy (DP) and privacy amplification techniques help optimize performance, reduce heterogeneity, and minimize leakage risks. For DeepJSCC+Encryption, device count impacts complexity and overhead, requiring distributed architectures, hardware acceleration, and efficient key management. The IBAL algorithm faces scalability issues, with higher computational complexity and training demands, where adversarial training mitigates reconstruction risks. For DeepJSCC+LBVQ, the number of devices affects codebook size and complexity, necessitating efficient management.  Moreover, the dynamic offloading framework plays a crucial role in large-scale privacy-preserving ToSC networks by effectively addressing dynamic traffic demands, optimizing resource utilization, and ensuring quality of service, thereby demonstrating significant application potential. For instance, the integration of traffic-aware network slicing and adaptive computation offloading strategies \cite{amin2025} facilitates precise traffic prediction and optimized resource allocation, thereby providing enhanced flexibility and adaptability in dynamic network environments \cite{cjm2025}. 

\subsection{Potential Applications}
These methods can optimize the transmission of semantically rich, compressed data while ensuring privacy in AR applications, especially in high-throughput, low-latency immersive environments. However, real-world AR deployment must tackle large-scale data, device heterogeneity, and dynamic content challenges. Privacy is also crucial in remote medical diagnostics with sensitive patient data. Evaluating system effectiveness using public datasets is a start, with future work focusing on low-latency, secure medical communication. For autonomous vehicles, real-time, reliable data exchange is vital. The proposed approaches can prioritize important data while preserving privacy, but network congestion and scalability issues need further study. Adapting to strict performance criteria and integrating with vehicle protocols like V2X is essential for this application.

\section{Future Research Directions}
Regarding securing privacy in ToSC, several research directions merit further exploration.
\subsection{Balancing Utility, Efficiency, and Privacy}
In ToSC systems, balancing utility, efficiency, and privacy presents a complex challenge. Utility denotes the system's effectiveness and efficiency in executing downstream tasks, ensuring tasks are completed accurately and promptly, thereby enhancing overall performance. Informativeness pertains to the volume of data transmitted, impacting transmission speed and bandwidth use. Security focuses on the system's ability to safeguard user privacy during data transmission and processing. While utility is paramount for transmitting semantically relevant data, its pursuit can inadvertently compromise privacy by exposing sensitive information. Efficiency is critical in 6G networks' high-volume, high-speed context, where conserving bandwidth and reducing latency are key to performance. Privacy, however, is the most difficult to balance. Although encryption offers robust security, it doesn't meet the demands of real-time, semantically-rich communication. More advanced strategies, despite their adaptability, introduce risks like the inexplicability of black-box deep learning. Achieving this balance typically requires a multi-layered approach \cite{10483054}, combining feature perturbation with intrinsic security techniques, and adapting to network and data dynamics. Regular audits and updates to privacy protocols are also vital to address emerging threats and technologies.

\subsection{Exploring Generative AI for Privacy Preserving}
Generative AI presents a novel approach to preserving privacy in ToSC. It focuses on creating data that is semantically similar to, but distinct from, the original dataset, thereby enabling the use of valuable data without exposing sensitive information. In ToSC, where the goal is to transmit semantically relevant information, generative AI can be used to produce high-quality synthetic data that maintains the statistical properties of the original dataset. This ensures that the utility of the data is not compromised, which is crucial for the effective functioning of ToSC systems. Additionally, since the synthetic data does not directly correspond to real user data, the risk of privacy breaches is significantly reduced. However, the use of generative AI in privacy preservation also poses challenges. One key issue is ensuring that the synthetic data does not retain any indirect identifiers that could lead to privacy breaches. This requires careful design and continuous evaluation of the generative models. Moreover, the computational complexity of training generative models can be a limiting factor.

\subsection{Transfer Learning for Task, Data and Channel Adaption} 
In JSCC-based ToSC, the system needs to be retrained as either the task, data or channel varies. Transfer learning can be instrumental for task/data/channel adaptation, allowing communication systems to efficiently adapt to new domains using pre-existing knowledge. Task adaptation focuses on applying learned models to new but related tasks. 
Transfer learning enables ToSC systems to quickly adjust to new tasks without the need for extensive retraining, thereby saving time and computational resources. Data adaptation is essential due to the variability of data types and sources. Transfer learning allows for the utilization of pre-trained models on one type of data (like text) and adapts them for different types (such as images or sensors data). This flexibility is particularly beneficial in multi-modal communication scenarios where different types of data need to be processed and transmitted seamlessly. Transfer learning can be also employed to adapt ToSC systems to varying channel conditions such as  interference, signal attenuation, and mobility. By learning from data transmitted under different channel conditions, a ToSC system can predict and adjust its parameters for optimal performance, even in less-than-ideal transmission environments \cite{amin2025}.

\section{Conclusion}
This paper highlights the significance of transmitting task-specific essential information efficiently while addressing the privacy preservation issue. It includes a comprehensive analysis of privacy-preserving strategies for ToSC, comparing feature perturbation methods like DP and encryption with intrinsic security incorporation approaches such as adversarial learning and LBVQ. Our research also explores experimental evaluations of these methods, assessing their performance and privacy protection capabilities. Finally, potential avenues for future study are provided.

\section*{Acknowledgments}
The work is supported in part by the National Natural Science Foundation of China under Grant 62171262 and 62301328; in part by Shandong Provincial Natural Science Foundation under Grant ZR2021YQ47; in part by the Taishan Young Scholar under Grant tsqn201909043; in part by Major Scientific and Technological Innovation Project of Shandong Province under Grant 2020CXGC010109.

\bibliographystyle{IEEEtran} 
\bibliography{bib}

\begin{thebibliography}{10}
\providecommand{\url}[1]{#1}
\csname url@samestyle\endcsname
\providecommand{\newblock}{\relax}
\providecommand{\bibinfo}[2]{#2}
\providecommand{\BIBentrySTDinterwordspacing}{\spaceskip=0pt\relax}
\providecommand{\BIBentryALTinterwordstretchfactor}{4}
\providecommand{\BIBentryALTinterwordspacing}{\spaceskip=\fontdimen2\font plus
\BIBentryALTinterwordstretchfactor\fontdimen3\font minus \fontdimen4\font\relax}
\providecommand{\BIBforeignlanguage}[2]{{%
\expandafter\ifx\csname l@#1\endcsname\relax
\typeout{** WARNING: IEEEtran.bst: No hyphenation pattern has been}%
\typeout{** loaded for the language `#1'. Using the pattern for}%
\typeout{** the default language instead.}%
\else
\language=\csname l@#1\endcsname
\fi
#2}}
\providecommand{\BIBdecl}{\relax}
\BIBdecl

\bibitem{8869705}
W.~Saad, M.~Bennis, and M.~Chen, ``A vision of 6{G} wireless systems: Applications, trends, technologies, and open research problems,'' \emph{IEEE Netw.}, vol.~34, no.~3, pp. 134--142, May 2020.

\bibitem{9759241}
C.~Feng, H.~H. Yang, D.~Hu, Z.~Zhao, T.~Q.~S. Quek, and G.~Min, ``Mobility-aware cluster federated learning in hierarchical wireless networks,'' \emph{IEEE Trans. Wireless Commun.}, vol.~21, no.~10, pp. 8441--8458, Oct. 2022.

\bibitem{10644029}
H.~Zhou, Y.~Deng, X.~Liu, N.~Pappas, and A.~Nallanathan, ``Goal-oriented semantic communications for {6G} networks,'' \emph{IEEE Int. Things Mag.}, vol.~7, no.~5, pp. 104--110, Aug. 2024.

\bibitem{10558819}
F.~Jiang, Y.~Peng, L.~Dong, K.~Wang, K.~Yang, C.~Pan, and X.~You, ``Large {AI} model-based semantic communications,'' \emph{IEEE Wireless Commun.}, vol.~31, no.~3, pp. 68--75, June 2024.

\bibitem{10670195}
F.~Jiang, L.~Dong, Y.~Peng, K.~Wang, K.~Yang, C.~Pan, and X.~You, ``Large {AI} model empowered multimodal semantic communications,'' \emph{IEEE Commun. Mag.}, vol.~63, no.~1, pp. 76--82, Jan. 2025.

\bibitem{10483054}
A.~Zhang, Y.~Wang, and S.~Guo, ``On the utility-informativeness-security trade-off in discrete task-oriented semantic communication,'' \emph{IEEE Commun. Lett.}, vol.~28, no.~6, pp. 1298--1302, Mar. 2024.

\bibitem{2020Adversarial}
T.~Xiao, Y.-H. Tsai, K.~Sohn, M.~Chandraker, and M.-H. Yang, ``Adversarial learning of privacy-preserving and task-oriented representations,'' in \emph{Proc. AAAI Conf. Artif. Intell. (AAAI)}, New York, USA, Feb. 2020, pp. 12\,434--12\,441.

\bibitem{10107616}
X.~Luo, Z.~Chen, M.~Tao, and F.~Yang, ``Encrypted semantic communication using adversarial training for privacy preserving,'' \emph{IEEE Commun. Lett.}, vol.~27, no.~6, pp. 1486--1490, June 2023.

\bibitem{10458014}
Y.~Wang, S.~Guo, Y.~Deng, H.~Zhang, and Y.~Fang, ``Privacy-preserving task-oriented semantic communications against model inversion attacks,'' \emph{IEEE Trans. Wireless Commun.}, vol.~23, no.~8, pp. 10\,150--10\,165, Mar. 2024.

\bibitem{9152658}
A.~Xiong, T.~Wang, N.~Li, and S.~Jha, ``Towards effective differential privacy communication for users’ data sharing decision and comprehension,'' in \emph{Proc. IEEE S\&P (SSP)}, San Francisco, USA, July 2020, pp. 392--410.

\bibitem{Shao2021LearningTC}
J.~Shao, Y.~Mao, and J.~Zhang, ``Learning task-oriented communication for edge inference: An information bottleneck approach,'' \emph{IEEE J. Sel. Areas Commun.}, vol.~40, pp. 197--211, Nov. 2022.

\bibitem{10159007}
S.~Xie, S.~Ma, M.~Ding, Y.~Shi, M.~Tang, and Y.~Wu, ``Robust information bottleneck for task-oriented communication with digital modulation,'' \emph{IEEE J. Sel. Areas Commun.}, vol.~41, no.~8, pp. 2577--2591, Aug. 2023.

\bibitem{Fredrikson2015ModelIA}
M.~Fredrikson, S.~Jha, and T.~Ristenpart, ``Model inversion attacks that exploit confidence information and basic countermeasures,'' in \emph{Proc. ACM. Conf. Computer. Commun. Secur. (CCS), Denver, CO, USA}, Oct. 2015, pp. 1322--1333.

\bibitem{amin2025}
A.~Mohajer, J.~Hajipour, and V.~C.~M. Leung, ``Dynamic offloading in mobile edge computing with traffic-aware network slicing and adaptive td3 strategy,'' \emph{IEEE Commun. Lett.}, vol.~29, no.~1, pp. 95--99, 2025.

\bibitem{cjm2025}
R.~Wang, Y.~Jing, C.~Gu, S.~He, and J.~Chen, ``End-to-end multitarget flexible job shop scheduling with deep reinforcement learning,'' \emph{IEEE Internet Things J.}, vol.~12, no.~4, pp. 4420--4434, 2025.

\end{thebibliography}

\newpage
\begin{IEEEbiographynophoto}
{Shuaishuai Guo} (shuaishuai\_guo@mail.sdu.edu.cn) received his B.E and Ph.D. degrees in communication and information systems from the School of Information Science and Engineering, Shandong University, Jinan, China, in 2011 and 2017, respectively. He visited the University of Tennessee at Chattanooga (UTC), USA, from 2016 to 2017. He worked as a postdoctoral research fellow at King Abdullah University of Science and Technology (KAUST), Saudi Arabia from 2017 to 2019. Now, he is a Full Professor of Shandong University. His research interests include 6G communications and AI Agent.
\end{IEEEbiographynophoto}

\begin{IEEEbiographynophoto}
{Anbang Zhang} (202234946@mail.sdu.edu.cn) is currently pursuing the M.S. degree with the School of Control Science and Engineering, Shandong University, Jinan, China. His research interests include semantic communications and machine learning.
\end{IEEEbiographynophoto}

\begin{IEEEbiographynophoto}
{Yanhu Wang} (yh-wang@mail.sdu.edu.cn) is currently pursuing the Ph.D. degree with the School of Control Science and Engineering, Shandong University, Jinan, China. His research interests include semantic communications and machine learning.
\end{IEEEbiographynophoto}

\begin{IEEEbiographynophoto}
{Chenyuan Feng} (Chenyuan.Feng@eurecom.fr) received the Ph.D.\ degree from Singapore University of Technology and Design in 2021. Currently, she is a research fellow at Eurecom, France and a Marie Skłodowska-Curie Scholar. Her research interests include edge intelligence and network native AI. Dr. Feng is also a receipt the 2021 IEEE ComComAp Best Paper Award and 2024 IEEE ICCT Best Paper Award. She serves as an Associate Editor for the IEEE Internet of Things Journal and the IEEE Open Journal of the Communications Society.
\end{IEEEbiographynophoto}

\begin{IEEEbiographynophoto}
{Tony Q. S. Quek} (tonyquek@sutd.edu.sg) received the Ph.D. degree from the Massachusetts Institute of Technology in 2008.  He is currently the Cheng Tsang Man Chair Professor with the Singapore University of Technology
and Design (SUTD) and an ST Engineering Distinguished Professor. He is also the Director of the Future Communications Research and Development Program (FCP), the Head of ISTD Pillar, and the Deputy Director of the SUTD-ZJU IDEA lab. He is a Fellow of IEEE, a Fellow of WWRF, and a Fellow of the Academy of Engineering Singapore, and the AI on RAN Working Group Chair in AI-RAN Alliance. 
\end{IEEEbiographynophoto}

\end{document}